# Complex interaction processes we need to visualize that successfully fill the quantum cup of a detector

Chandrasekhar Roychoudhuri[a], Narasimha S. Prasad[b],
[a]Physics Department, University of Connecticut, Storrs, CT 06268, USA
[b]NASA Langley Research Center, 5 N. Dryden St., MS 468, Hampton VA, 23681, USA

## ABSTRACT

Sensors are measuring tools. In any measurement, we have at least two different kinds of interactants. We never know all there are to know about any one of these interactants and the interaction processes that are mostly invisible. Yet, our engineering innovation driven evolution is persisting for over five million years. It is then important to articulate explicitly our Interaction Process Mapping Thinking (IPM-T) that we keep applying in the real world without formally recognizing it. We present how the systematic application of IPM-T removes century old wave-particle duality by introducing a model of hybrid photon. It seamlessly bridges the quantum and the classical worlds. Photons are discrete energy packets only at the moment of emission; then they evolve diffractively and propagate as classical waves. Thus, "interference of a single indivisible photon" is only a non-causal assertion. We apply IPM-T to improve the photoelectric equation & we obtain Non-Interaction of Wave (NIW) amplitudes. Note that Huygens explicitly articulated NIW by postulating his "secondary spherical wavelets". Later Fresnel incorporated this postulate in the now famous Huygens-Fresnel (HF) diffraction integral. Most modern optical science and engineering are based upon propagating EM waves through optical devices and systems using this integral in some form or other. Maxwell's wave equation accepts HF integral as its solution. Systematic application of IPM-T to our causal and working mathematical equations, along with NIW in interferometric experiments, reveal that Superposition Effects can emerge only when the interacting material dipoles respond, whether classically or quantum mechanically, to the joint stimulations due to all the simultaneously superposed waves. This indicates the non-causality of our belief that a single indivisible photon can interfere by itself. We would not have a causally evolving universe had any stable elementary particle were to change itself through self-interference. Further, our working superposition equations always contain two or more terms representing two or more independently evolving entities. That is why we need physical instruments with two or more independent channels of propagation along with appropriately placed detectors to generate physical superposition effects. Nature does not violate causality. Otherwise, our causally framed equations would not have been working so elegantly.

**Keywords:** Interaction processes in QM; Semi-classical model; Hybrid photon model; Non-Interaction of Waves or NIW; Proper characterization of light & detecting molecules; Causality of Superposition Effect; Un-observability of optical Superposition Principle.

## 1. INTRODUCTION

Optical engineers never analytically propagate isolated indivisible light quanta since QM formalism has never provided one. We still use, quite successfully, Huygens-Fresnel Diffraction integral (or its variations) and Maxwell's equations to solve all optical propagation related engineering problems. Classical Optics has been modeling nature correctly for well over two hundred years. Growth of optical sciences have been continuing unabated. This is true even for the rapidly growing fields of Nano photonics and Plasmonic photonics.

Let us underscore that the formalism of Quantum Mechanics (QM) also has been working. Current QM formalism correctly predicts that the energy transfer has to be $h\nu$. All quantum detectors need to fill up their "quantum cups" with precisely $h\nu$ quantity of energy out of the Maxwell's wave packets while the dipole elements in the detector is undergoing quantum level or band transition. The required quantum amount of energy can also be provided through kinetic collision with another particle provided its kinetic energy, $(1/2)m\mathrm{v}^2 \geq h\nu$. Schrodinger's and Heisenberg's mathematical formalisms (since 1925) have been successfully modeling atoms and molecules, specifically, light emission and absorption. Quantum Optics has been modeling nature for 94+ years. However, QM does not give us any causal or rational description of the physical processes taking place in between the initial to final quantum transition. This is the key and persisting weakness of QM formalism. We bridge this gap by presenting a model for a hybrid photon – a quantum of energy $h\nu$ gets released by an excited atom, which then immediately evolves into a Maxwellian time-finite wave packet.

Note that Einstein's "quantum" of 1905 postulate was framed (i) eight years before Bohr's quantum atom (quantized electron), and (ii) twenty years before the QM formalism were developed. Had Einstein assigned the "quantumness", which he correctly observed in the photoelectron data, to the electrons, instead of to the classical EM waves, he would have been the father of Quantum Mechanics. Even though Einstein postulated that light energy has to be quantized ("indivisible light quanta"), the QM formalism does not require that a quantum transition can be triggered exclusively by another quantum entity that has the exact quantum of energy hv to donate. Atoms and molecules, to fill up their quantum cups with the required energy $h\nu$, routinely and selectively absorb classical kinetic energy (heat) from the kinetic particles. This is why Boltzmann's classical statistical law of temperature dependent population-density works so well for quantum mechanical ensemble in thermal contact with each other. A much simpler example is a He-Ne laser-tube. Classical, free kinetic electrons, accelerated by high voltage potential difference, collide with the He and Ne atoms; shares the necessary amount of energy hv to fill up the quantum cups of He and Ne atoms out of their total kinetic energy, $(1/2)m\mathrm{v}^2$. The atoms are raised to their upper-quantum lasing levels. After reaching the anode, the classical electrons are recycled externally as a classical electric current.

The proposed hybrid photon [1,2] contains a quantum of energy $h\nu$ at the moment of emission out of atoms and molecules, which immediately evolves into a classical wave packet and follows classical diffraction integral and Maxwell's wave equation. Our quantum detectors (the stimulated dipole molecules) fill up their "quantum cups" [3] when there is sufficient frequency-resonant energy density in the incident classical field. Recognition of this semi-classical model will open up much deeper understanding of light-matter interaction processes [4], especially, at extremely low light levels. The appearance of dark fringes should be correctly recognized as due to the "local" resultant E-vector being zero, the detecting molecule are not stimulated and hence cannot absorb any energy out of the passing-by fields [5]. It is not due to the "non-arrival" of any indivisible light quanta. One can also appreciate the universal Non-Interaction of Waves (NIW) [1], which has been formally proposed by Huygens in his book, "Treatise on Light" [6]. The un-necessary postulate of "wave-particle duality" can also be finally erased out of our literature.

The paper will also re-visit the concept of "single photon interference". If it is a correct natural phenomenon, then it should be valid for all possible superposition effects with light beams. Consider the well-established LIDAR technology. Here, one sends out a laser beam to be scattered back from some remote air molecules with a Doppler frequency shift. The returned signal is then combined with a local laser oscillator on a high speed photo detection system, which then discerns the heterodyne (time-varying) beat signal (frequency difference between the local oscillator and the returned signal from the distant moving air molecules) [7]. Mathematically, it is a simple two-term amplitude-frequency superposition equation, like any other two-beam interference experiments. Can we really create a causally valid story that the time-varying signal is really generated by "indivisible single photons" at a time, reaching the detector one at a time? With the well-understood semi-classical model [8-12], the explanation is trivially obvious. The ensemble of detecting dipoles, under the influence of two-different phase-steady E-vector frequencies, undergo time-varying dipolar stimulation. The ensemble of detecting dipoles keep filling up their "quantum cups" with the time-varying energy available from all the superposed fields out of both the superposed beams. Accordingly, they keep releasing electrons whose number is time-varying and proportional to the square modulus of the sum of the resultant dipolar stimulation induced by the two classical wave amplitudes with different frequencies.

The paper is structured into following four sections.

Section-2: Will briefly present Interaction Process Mapping Thinking (IPM_T). This is an efficient engineering approach to help articulate the generic interaction process steps behind all data-gathering experiments. This approach is applicable to all phenomena. This will also help us to seek out ontological reality of nature.

Section-3: We will re-frame "duality" through nature's causal functional behavior to articulate what is a hybrid photon that bridges both classical and quantum observations. Both the quantized emission and absorption will be accommodated by defining "transient photons" released by atoms, immediately evolves into classical wave packets that propagate out diffractively following Huygens-Fresnel diffraction integral and/or Maxwell's wave equation.

Section-4: We will differentiate the measurable *quadratic* Superposition Effects (SE) from the non-measurable *linear* Superposition Principle (SP). We will use demonstrated experiments to establish that causal and mathematical superposition equation imply real physical superposition of more than two signal simultaneously interacting with some isolated or an assembly of detecting dipoles. The readers will then recognize that most superposition effects (re-organization of wave energy) happens as classical boundary value problem.

Section-5: This section concludes the paper by presenting the technical summary and then articulates the significant value of the paper.

## 2. INTERACTION PROCESS MAPPING THINKING – AN EFFICIENT ENGINEERING APPROACH

The concept of wave-particle duality started with Newton ("Corpuscular") and Huygens ("Secondary Wavelet") around the second half of 1600. Both Newton and Huygens recognized that their "duality" debate represented their ignorance about the deeper knowledge of light. However, during the 1900, we slowly converted the "wave-particle duality" into a new knowledge without discovering the real structural difference between waves and particles. This why the debate over the wave-particle-duality continues even though it is being muzzled continuously. However, the muzzling only slows down our collective deeper enquiry in finding the ultimate resolution – the causal reality, or ontological reality of nature, which the Copenhagen Interpretation implies not to be achievable by human theories. Surprisingly, Newton and Huygens were more hopeful. Of course, if we do not enquire, we will never find out.

In this paper, we are trying to articulate how to go beyond Copenhagen Philosophy by recognizing the inherent and natural limitations behind developing the evidence based science and how to overcome them iteratively in steps. First, we need to recognize that all of our scientific measurement processes suffer from the inherent Problem of Information Retrieval, or PIR. The "Measurement Problem" is not something that can be overcome by elegant mathematical theorems. Human invented mathematical logics do not have direct access to nature's interaction process maps. This is why our evidence-based science cannot develop any "final" theory. We do not know in complete and exhaustive details pertaining to any entity in this universe. Our engineers must keep providing guidance through newer and newer innovations to extract more and more information out of nature. We still do not know whether we can ever design instruments that will have consistent fidelity of 100% accuracy. However, our most challenging and insurmountable problem is that we can never gather all the information about anything through any set of experiment. We do not know complete information about any of the interactants we chose for an experiment. Further, the details of most of the interaction processes in the micro world are beyond our direct visualization. Fortunately, the steady successes of scientific and engineering activities, using causal mathematical approach, clearly imply that the rules (or the cosmic logics) behind *interaction processes* in nature are invariant. Therefore, understanding and visualizing (through rational imaginations) nature's interaction processes represent a firm platform to anchor our thinking. We must keep advancing our working theories iteratively in steps, while challenging our working theories and re-build, re-structure them in repetitive incremental step. The following summary will help us appreciate the significance of the Problem of Information Retrieval [13].

(i) Measurable data represent some quantitative physical transformations in interactants.
(ii) Physical transformations must be preceded by energy exchange.
(iii) Energy exchange must be guided by some force of interaction.
(iv) The mutual force of interaction must be actually experienced by the interactants.
(v) The experience of the force of interaction is effectively the "Physical Superposition" of the interactants, since all forces are decay with distance.
(vi) All physical transformations due to energy exchange require finite durations as their centers are always at finite distance from each other.
(vii) Corollary: Impossibility of force-of-interaction-free transformation, which demands re-visiting the postulate of "entanglement".

## 3. RE-FRAME "DUALITY" THROUGH NATURE'S CAUSAL AND FUNCTIONAL BEHAVIOR

QM is correct that atoms and molecules emit and absorb discrete amount of quantized energy. Classical optics is also correct that we have been consistently advancing using Huygens-Fresnel diffraction integral (or its variations) and Maxwell's wave equation. Then the two theories must be bridge-able using appropriate causal postulate and using the Interaction Process Mapping Thinking (IPM-T), as outlined in the last section. In this section, we develop how to accommodate both quantized emission by quantum entities and also absorption of a quantum "cupful" of energy out of light waves, propagating always as classical Maxwellian wave packets.

### 3.1. Accommodating quantized emission

Quantum formalism is correct that the energy released in any quantum transition is always uniquely quantified as $\Delta E_{mn} = h\nu_{mn}$. Therefore, we postulate that a "transient photon" of energy $h\nu_{mn}$ is released during a downward transition containing the characteristic frequency information $\nu_{mn}$. Then, immediately after release, this transient photon evolves into a quasi-exponential temporal wave packet, with a carrier frequency of $\nu_{mn}$ for the E-vector oscillation, which obeys Maxwell's wave equation as it propagates out. The temporal shape is the wave packet is quasi-exponential containing the total energy $\Delta E_{mn}$. The quasi-exponential shape is chosen to preserve the causality that the signal starts from zero, rises to a final peak value quite rapidly and then dies out exponentially to zero. The dominant exponential shape preserves the predicted and observed natural line width of Lorentzian shape. Note that classical spectrometers give rise to a measured linewidth that is the Fourier transform of the temporal envelope function [1, see also Ch.5 and Ch.10 in ref.2]. The sketch in Figure 1 illustrates this model.

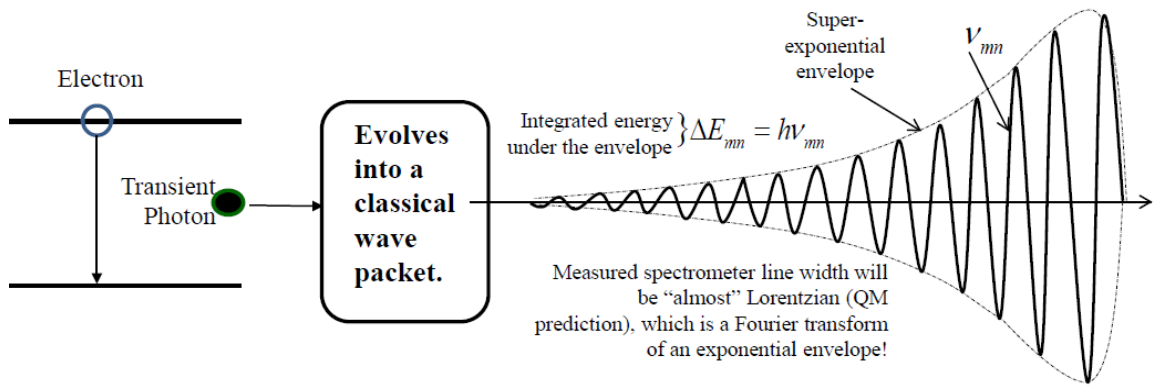

*Figure 1. Accommodating quantized emission from excited quantum entities. A "transient photon" of allowed energy $h\nu$ is released by a quantum entity, which immediately evolves into a quasi-exponential wave packet carrying the energy $\Delta E_{mn} = h\nu_{mn}$ with the required carrier frequency $\nu_{mn}$. A classical spectrometer will respond to this quasi-exponential envelope by generating a Lorentzian linewidth, corroborating experimental results.*

### 3.2. Accommodating quantized absorption out of multiple classical wave packets

All EM waves propagate through diffractive spreading. Therefore, the total energy $\Delta E_{mn} = h\nu_{mn}$ of an expanding wave packet can no longer be available within one-cubic-Angstrom atomic volume for direct absorption by an atom, unless the emitting and absorbing atoms are confined within a micro-confocal cavity [14]. However, in most situations, the source of light and the light absorbing detecting molecule are widely separated in space. Diffraction thins out the energy density of all emitted wave packets. The only way for an atom to collect the required $h\nu_{mn}$ quantity of energy is to collect it out of a large number of diffracted wave packets. We present this model in the sketch of Figure 2 (the left diagram).

We need to incorporate another set of postulates. Once stimulated by a resonant EM wave frequency, the effective atomic dipolar interaction cross-section becomes enormously large, many wavelength square. This could be thought of as that the resonant atomic dipole as *pulling* in the wave energy out of a very large frontal (oncoming wave) volume. This effect is further enhanced by the wave itself, which tries to converge (*push*) its energy on to the stimulated dipole. The rationale behind the introduction of this *push-pull* postulate is supported by, first, (i) from the theoretical modeling of radiation field converging on a resonant dipole (see Fig.2, right diagram); second, (ii) from the observations in many fast and efficient resonance fluorescence experiments, which is quantified in Fig.3; third, (iii) from the observed transition time period for photoelectron release is always below a femtosecond [15]. We summarize this collective behavior of resonantly excited atoms as behaving like a large *quantum cup,* which absorb the necessary quantum of energy through co-operative *push-pull* mechanism. This is further illustrated in the section below.

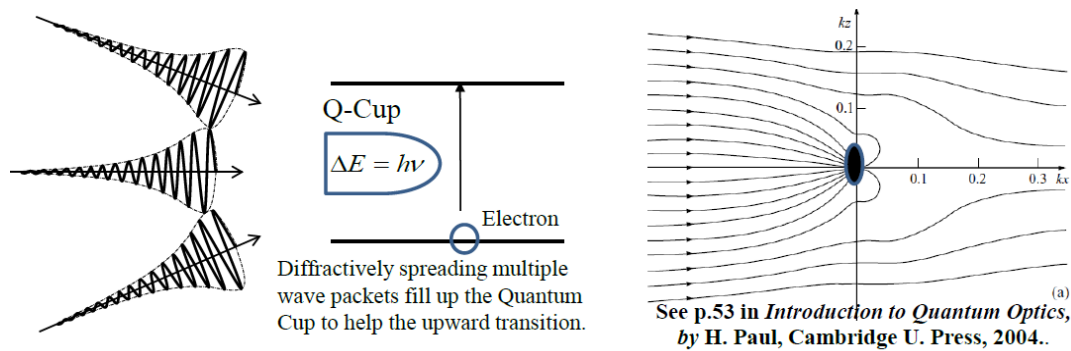

*Figure 2. Left: Accommodating quantized absorption of energy out of multiple, energy-divisible, classical wave packets. Right: Showing how the "field lines of force" converge on a small, resonetically stimulated dipole.*

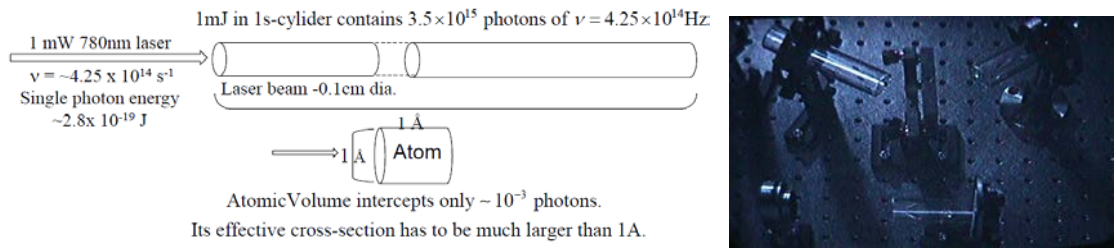

*Figure 3. **Left sketch:** Computation of the number of red photons that could exist within the volume of one-Angstrom-cube of a typical atom, which is a very small number given a 1mW 780nm laser beam of 1mm cross-section. **Right photo:** Rb-fluorescence easily observable along the stimulating laser beam. A resonant atom must have a much larger effective cross-section for interaction with light compared to its actual physical size[7].*

### 3.3. Experimental validation of atom's large "Quantum Cup" Suction volume!

Resonance fluorescence experiments with low-pressure gas atoms, like Rb, illuminated by a frequency-tunable laser beam, is well known technique for spectrometric studies of various quantum properties of atoms [16]. During one such experiment in our lab [7] we observed that a one mW 780nm laser beam of diameter 1mm can instantaneously generate the resonance fluorescence of spontaneous emission visible to the naked eye in the ambient light of the laboratory. A simple calculation shows that there could be only about ~$1.2 \times 10^{-17}$ red-photon within a volume of one-Angstrom-cube of any Rb-atom (see Fig.3, left sketch). Therefore, the probability of stimulated absorption of a photon and then spontaneous re-emission should have been significantly weaker than what we had observed along the entire Gaussian cross-section of the laser beam (see Fig.3, right photo). We present these observations as semi-quantitative validation that a stimulated atomic dipole functions like a large

"quantum cup" that can absorb energy through push-pull mechanism out of a large volume containing multiple EM wave packets.

### 3.4. Re-visiting the photo-electric equation from the stand point of semi-classical model

Einstein's brilliance was that he was the first one to notice the quantumness in the photoelectric data with the optical frequency of the illuminating light (see Fig.4, left curve). Therefore, Einstein applied the prevailing philosophy of Measurable Data Modeling Thinking (MDM-T), and presented the energy balancing equation, heuristically assigning the "quantumness" to classical EM waves.

$$h\nu = \phi_{work\ fn.} + (1/2)m v_{el.}^2 \tag{1}$$

Had Einstein applied concurrently the Interaction Process Mapping Thinking (IPM-T), perhaps, he would have been guided to think about the origin of the "work function" and the consequent necessity of quantum mechanically stimulating the electron-nucleus bound system within the material first. The concept of frequency-resonant amplitude stimulation of electron-nucleus bound system would have become obvious. In that case, Einstein would have invented the quantum mechanics, perhaps, with a different mathematical formalism approach compared to the prevailing formalism. Einstein did say towards the end of life that in spite of 50 years' of brooding, he was still confused about the exact nature of light. However, he did not deviate from his old enquiring question, "what are light quanta?"

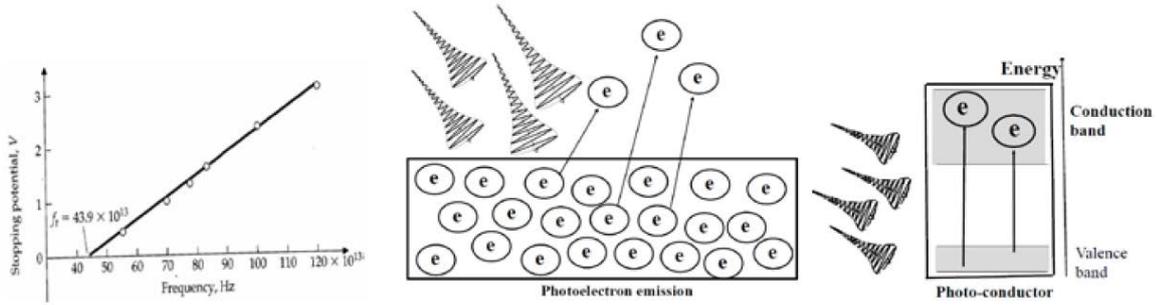

*Figure 4. Re-visiting the photo-electric emission from the stand point of semi-classical model.* **Left curve** *reveals the quantumness in optical frequency [17]. The right two sketches pictorially depicts electron release, or electron transfer due to photo excitations induced by multiple classical wave packets.*

Based on our current knowledge of QM and semi-classical model [8-10, 18,19], we present a non-rigorous formulation. The elementary stimulation $\psi$ of the dipole-complex holding the electron can be written, using the linear dipolar stimulation parameter, $\chi(\nu)$ as:

$$\psi = \chi(\nu)E(\nu) \tag{2}$$

In reality, we have to take into account of the amplitude stimulations induced by multiple diffractively spreading wave packets, which eventually help fill up the quantum-cup of the dipole-complex (see Fig.4). Then the energy transfer process requires the square-modulus step. Accordingly, we write:

$$\psi_{res.} = \sum_q \chi(\nu_q)E(\nu_q) \tag{3}$$

$$|\psi_{res.}|^2 = \left|\sum_q \chi(\nu_q)E(\nu_q)\right|^2 \tag{4}$$

Plots of data for quantum events are plotted as relevant only by taking ensemble average:

$$\left\langle|\psi_{res.}|^2\right\rangle = \left\langle\left|\sum_q \chi(\nu_q)E(\nu_q)\right|^2\right\rangle \Rightarrow \left\langle h\nu_q \right\rangle_{Many\ QM\ Cups} = \left\langle \phi_{work\ fn.} + (1/2)m v_{el.}^2 \right\rangle \tag{5}$$

The left hand equality of Eq.5 represents the ensemble average of electron release. The right hand equality shows that, under ensemble average, Einstein's direct "energy balance equation" becomes equivalent to the current quantum formalism. However, the left hand equation captures the two-step physical processes that take place before the electrons are "released". Again, the two steps are: (i) The dipolar stimulation of the bound electrons; followed by (ii) quantum-cupful of energy absorption by the dipole-complexes to "release" the electrons. This is again an explicit application of the Interaction Process Mapping Thinking (IPM-T).

### 3.5. Discovering Non-Interaction of Waves by imposing IPM-T on the photoelectric equation

To strengthen the significance of IPM-T, let us now extract more physics out of a *working* mathematical equation by scrutinizing the physical properties of the mathematical symbols and the operators. We will discover Non-Interaction of Waves (NIW) [2]. Let us consider Eq.4. As per standard algebraic rulel, $\chi(\nu_q)$ cannot be taken out of the summation operator, of the summation sign, since the product, $\chi(\nu_q)E(\nu_q)$, is interdependent through the variable $\nu_q$. The strength of the responsivity of all detectors, as dipoles, is frequency dependent.

$$|\psi_{res.}|^2 = \left|\chi_\nu \sum_q E(\nu_q)\right|^2 = \chi_\nu^2 \left|\sum_q E(\nu_q)\right|^2 \qquad (6)$$

However, if we assume that we are using a filtered narrow-line beam of light, then $\chi$ becomes a constant for the given frequency. Then we are allowed to take out the polarizability $\chi_\nu$ and $\chi_\nu^2$ out of the summation operator. Let us now compare the meanings implied by Eq.4 and Eq.6. The Eq.4 implies that the summation operation is carried out by the same dipole-complex experiencing all the stimulations $\sum_q \chi(\nu_q)E(\nu_q)$ simultaneously. The operator is the dipole-complex. However, once we take the polarizability factors, $\chi_\nu$ and $\chi_\nu^2$, out of the summation-operators, as presented in the two steps of Eq.6, the physical implication is dramatically different. The summation operation $\sum_q E(\nu_q)$ implicates that different EM wave amplitudes, when physically superposed, can execute a linear operation of summing themselves and reorganize their mutual amplitude distribution. Similarly, $\left|\sum_q E(\nu_q)\right|^2$ implies that, besides re-organizing their mutual amplitudes, superposed EM waves can also execute the non-linear square modulus operation by themselves to re-organize their mutual energy distribution. However, all propagating waves, being linear excited states of some parent tension field, simply co-propagate or cross propagate without affecting each other's amplitudes or energies. This is Non-Interaction of Waves (NIW), which we have been neglecting to explicitly recognize in all branches of physics for centuries [see Ch.2 in ref.2]. The reason is that we have been neglecting to incorporate IPM-T over and above our very successful Measurable data Modeling Thinking (MDM-E). In other words, we have been failing to enquire the about the physical (ontological) interaction processes in nature that give rise to measurable data in our instruments.

## 4. DIFFERENTIATING MEASURABLE SUPERPOSITION EFFECT (SE) FROM NON-MEASURABLE SUPERPOSITION PRINCIPLE (SP)

In this section, we will underscore that the *linear* mathematical Superposition Principle (SP) in optics is not an observable. Only the Superposition Effect (SE) can become observable after the execution of the nonlinear square modulus operation of the stimulations the materials and detectors experience due to all the superposed fields on them. The previous section 3.5 has already established the related reasons using mathematical logics and IPM-T. In this section, we will re-produce some selected experiments from previously published paper by the author [20] to experimentally establish the importance of differentiating between SP vs.SE. Our discussions will also reveal that most observable SE, or superposition fringes as energy re-distribution or re-direction, become manifest due to classical boundary value problems. Here are some obvious examples. Generation of patterned amplitude distributions, as eigen modes of the diffraction integral, out of enclosed optical cavities or channeled propagation through very long single mode fibers. No QM concepts have ever been needed to derive and/or understand emergence of spatial modes. Quantum properties become relevant, as in photoelectric effect, only at the final stage when the detectors are clearly quantum mechanical. It is correct that all final *optical* detectors are quantum mechanical. This is true even for human vision. The retinal molecules must first undergo electronic transition due to the incident light, which triggers the final ion-diffusion process to generate the "visual signals" as potential differences.

**4.1. Mach-Zehnder interferometer (MZI) with collinear Poynting vectors; a classical superposition effect**

The left sketch in Fig.5 shows an MZI in *scanning mode* that generates temporally oscillating fringes when one of the MZI mirror is being scanned. Two beams, produced by the beam splitter BS are aligned and combined by the beam combiner BC in such a way that the two output pairs of beams, heading towards the detectors D1 and D2, have perfectly collinear and coincident Poynting vectors. Under these conditions, we can observe the fringes due to superposition effect only as output energy oscillation as the mirror M1 is scanned using a PZT driven by a saw-tooth wave. The middle photo shows the energy received by the detectors D1 and D2. They are inverse sinusoids. The energy of the two ***steady and equal*** input amplitudes are emerging as complementary and sinusoidally *oscillating beams of energy between zero and one*. This is physical re-direction of energy from one beam to the other. Effectively, the values of the reflectance $R$ and the transmittance $T$ are both dynamically oscillating between zero and one as the incident phase values of the two beams keep oscillating, even though the dielectric coating is 0.5/0.5, fixed by design. This is a pure classical boundary value behavior of the collective surface dipoles. It is critical to appreciate that the energy re-direction is engendered due to the $\pi$ phase shift in the ***external reflection*** [21] (see thick rays in the right sketch of Fig.5). When the physical path delay between the two beams superposed on the BC is exactly modulo-$2\pi$, the reflected right-going beam suffers a relative $\pi$ phase shift. The *right-going transmitted beam* is now out-of-phase with the *right-going reflected beam*. The active boundary dipoles cannot emit light in the right-going direction. Under this condition, all the energy due to both the incident beams travels along the up-going direction. This has nothing to do with quantum mechanics. There is no quantum physics or quantum transition behind this physical process of energy-redirecting capability of a 0.5/0.5 beam splitter, whose effective $R$ & $T$ values dynamically oscillate between complementary zero and one.

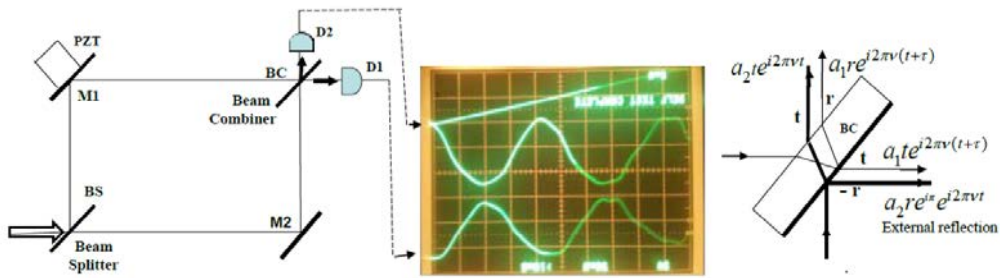

*Figure 5. A Mach-Zehnder interferometer (MZI) with output beams having perfectly collinear Poynting vectors. The boundary conditions determine which way the energy will be directed by the beam combiner (BC)* **Left sketch:** *The MZI in scanning mode.* **Middle photo:** *The two sinusoidally oscillating output intensities, displayed on a two channel scope. The total energy remains conserved. Both the boundary values of the beam combiner, the reflectance and the transmittance, oscillate between zero and one, purely based on the effective phase conditions of the two beams incident on the opposite sides of the BC.* **The right sketch:** *The critical enabler of this classical superposition effect is the π-phase shift of the beam undergoing external reflection (thick ray).*

Let us frame the mathematical model using classical superposition principle, while incorporating the linear polarizability parameter $\chi$ to represent the dipolar stimulation of the boundary layer. Here $\chi$ represents the bulk boundary polarizability of the beam splitter molecules. The active polarizability role of the boundary layer is at the root of Brewster's Law, Malus' law and partial polarization of "un-polarized" light in reflection. EM waves always stimulate the material dipoles to oscillate along the direction of its E-vector while the wave propagates through any medium [21]. The dipoles in the boundary layer oscillates preferentially along the surface. Let us assume $A_{Rt.}(t,\tau)$ represents the sum of the two right-going amplitudes and $A_{Up}(t,\tau)$ represents the two up-going amplitudes.

$$A_{Rt.}(t,\tau) \equiv \chi(\nu)E(t,\tau) = \chi(\nu)E_1(t+\tau) + \chi(\nu)E_2(t)$$
$$= \{\chi(\nu)a_1 t\}e^{i2\pi\nu(t+\tau)} + \{\chi(\nu)a_2 r e^{i\pi}\}e^{i2\pi\nu t} \quad (7)$$
$$= \chi_\nu[a_1 t e^{i2\pi\nu(t+\tau)} - a_2 r e^{i2\pi\nu t}]; \text{ frequency band very narrow.}$$

$$A_{Up}(t,\tau) = \chi_\nu[a_1 r e^{i2\pi\nu(t+\tau)} + a_2 t e^{i2\pi\nu t}]; \text{ frequency band very narrow.} \quad (8)$$

The detectable intensities in the right-going and up-going beams are now given by the Eq.9 and Eq.10, respectively. For mathematical simplicity, we are assuming that the dielectric boundary layer is approximately loss-less.

$$I_{Rt.}(\tau) \equiv |\psi_{Rt.}(\tau)|^2 = \left|[a_1 t e^{i2\pi\nu(t+\tau)} + a_2 r e^{i\pi} e^{i2\pi\nu t}]\right|^2$$
$$= [(a_1^2 t^2 + a_2^2 r^2) - 2a_1 a_2 tr \cos 2\pi\nu\tau] \quad (9)$$
$$= a^2[1 - \cos 2\pi\nu\tau]; \text{ only if } r^2 = t^2 = 0.5 \text{ and } a_1^2 = a_2^2 = a^2$$

$$I_{Up.}(\tau) \equiv |\psi_{Up}(\tau)|^2 = \left|a_1 r e^{i2\pi\nu(t+\tau)} + a_2 t e^{i2\pi\nu t}\right|^2$$
$$= [(a_1^2 r^2 + a_2^2 t^2) + 2a_1 a_2 tr \cos 2\pi\nu\tau] \quad (10)$$
$$= a^2[1 + \cos 2\pi\nu\tau]; \text{ only if } r^2 = t^2 = 0.5 \text{ and } a_1^2 = a_2^2 = a^2$$

Summing the second lines of the above two equation we can appreciate the generic law of conservation of energy for the MZI in *scanning mode*, when $a_1 \neq a_2$:

$$D_{Rt..}(\tau) + D_{Up.}(\tau) = (a_1^2 t^2 + a_2^2 r^2) + (a_1^2 r^2 + a_2^2 t^2)$$
$$= (t^2 + r^2)(a_1^2 + a_2^2) = (a_1^2 + a_2^2); \text{ for loss-less beam splitter} \quad (11)$$

Under the conditions of equal amplitudes arriving on the BC having 0.5/0.5 coating, as expressed in the last lines of Eq.9 and Eq.10, the sum total energy in the two beams remains conserved:

$$D_{Rt..}(\tau) + D_{Up.}(\tau) = 2a^2, \text{ with } a_1^2 = a_2^2 = a^2; \text{ for loss-less beam splitter} \quad (12)$$

It is important to notice that in the above Eq.7 through Eq.12, to utilize the causal physical process of re-directing energy out of one beam to the other, we must have both the beams physically present simultaneously. *Otherwise, these equations would not represent causal physics*. Therefore, even if single photons existed, we would need two of them to be simultaneously present from the opposite sides of the beam combiner to generate the physical conditions to re-direct themselves along one way or the other! If one meticulously aligns the interferometer to send 100% of the energy, say, in the right-going channel and 0% in the upward-channel, just by introducing "photon counting" electronics, one cannot argue that they have "indivisible light quanta"! Observable superposition effect is the mundane classical response of the beam splitter boundary under the conditions already defined above.

### 4.2. Mach-Zehnder interferometer with non-collinear Poynting vectors; classical and/or quantum superposition effect!

When the Poynting vectors of the output beams are non-collinear, the beam combiner BC continues to behave just as the beam splitter BS (see Fig.6, left sketch). The $R \& T$ values remain fixed and keeps function as manufactured. The stimulations of boundary dipole complexes due to the two non-collinear electric vectors are independent of each other. The superposition fringes are now fixed in space; the spatial frequency of which is determined by the angle between the two wavefront-vectors (see the middle sketch of Fig.6). The fringes as spatial intensity variations can be visually observed, or recorded by a camera focused on the ground-glass surface. the fine pits and lumps of the Silica surface, when smaller than the wave length of the light, respond to the local resultant electric- vector stimulation. Thus, at whichever space point the resultant electric-vector-stimulation is zero. The pits and lumps cannot generate forward scattered light. Accordingly, these locations appear to the eye or the camera, as dark fringes. Similarly, all space points on the ground glass that receive above-zero resultant E=-vector stimulations; generate forward scattering of light, which will appear to the eyes, or the camera, as bright fringes. The Silica lumps and pits are large classical objects and the optical scattering phenomena and we model this phenomenon quantitatively using classical formalisms [22]. Of course, we can chose to record these fringes with a quantum mechanical device like a very sensitive CCD camera and start counting "clicks" it generates at reduced light level. The CCD signal-displaying elements are spatially discrete. Therefore, at very weak levels of light, the spatially discrete CCD elements will keep integrating charges randomly at different elements, since the charge generation physics is quantum mechanically a random phenomenon. Therefore, we should resist applying the quantumness in quantum detectors as that of light. Einstein had made the same mistake some 113 years ago when the quantum mechanical binding of electrons in all materials were not understood.

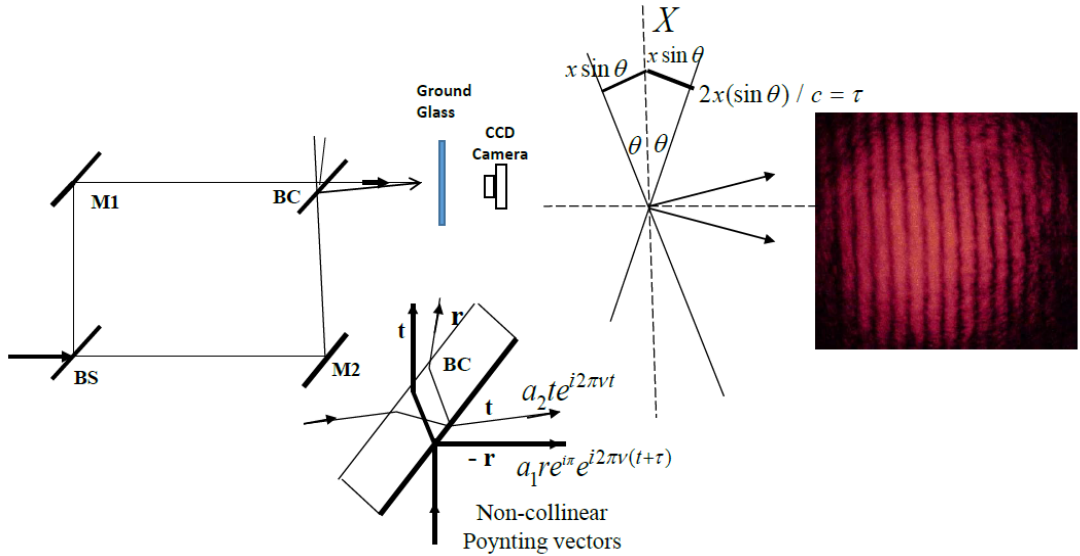

*Figure 6. Non-collinear output beams from an MZI can generate only spatially varying fringes. Even these fringes can be made observable by employing the classical phenomenon of light scattering and using a glass scatter-plate. Quantum devices like CCD camera can also register the fringes. However, the success of a quantum device to record superposition fringes does not make the superposition phenomenon as exclusively belonging to a quantum phenomenon.*

The superposition relation for the right-going beams can be expressed as:

$$A(\tau) = \chi_\nu [a_1 R^{1/2} e^{i\pi} e^{i2\pi\nu(t+\tau)} + a_2 T^{1/2} e^{i2\pi\nu t}] \qquad (13)$$

We have assumed that the incident radiation is a single frequency laser beam and hence $\chi_\nu$ is a constant. This polarizability will now represent for any material that is used to make the fringes become visible, the ground glass or the CCD-camera material. Fringe intensity now varies spatially with $\tau = 2x\sin\theta / c$, which is defined in the middle sketch of Fig.6.

## SUMMARY & THE VALUE OF THE PAPER

### 5.1. The Summary

The wave-particle duality has been with us as a serious debate since it began with Newton (*corpuscular*) and Huygens (*secondary wavelets*). The duality debate has not been resolved with proper new knowledge about differentiating waves and particles. We still do not know the precise structural knowledge about electrons and photons. However, our engineers have figured out how to (i) generate, (ii) modulate, (iii) propagate and (iv) detect electrons and photons. These engineers have facilitated the onset of the Knowledge Age. Accordingly, we have started by articulating the engineering mode of thinking – Interaction Process Mapping Thinking (IPM-T). The rest of the paper demonstrates how to use IPM-T to remove wave-particle duality for light waves.

First we presented a hybrid photon model that seamlessly bridges the perceived gap between quantum and classical descript ion of EM waves. Excited quantum entities do release discrete quantum of energy $\Delta E_{mn} = h\nu_{mn}$; but we have posited that this quantum of energy immediately evolves into a classical wave packet and propagates following classical diffraction integral that has not been violated for more than two centuries. The "bridging the gap" has also required us to propose that during the absorption of radiation energy, the quantum dipoles acts as a larger quantum cup and the energy transfer process is a collaborative one "push" by the field and "pull" by the excited dipole. This accounts for how atoms absorb the desired amount of energy out of a spatially large field of propagating waves.

Then we have applied the above concept to reframe Einstein's photoelectric equation using our semi-classical concepts summarized above.

We have then focused on logically and experimentally demonstrating that the concept of "single photon interference" is logically inconsistent with our working and causal equations of superposition. Further, systematic application of IPM-T clearly shows that the superposition effects can observed most of the time using pure classical optics, without the need of any quantum detectors. Our experiments also underscores the physical necessity of the simultaneous presence both the superposed signals on the interference-effect generating material. Thus, "single indivisible photon interference" is a non-causal proposition.

**5.2. The significance of the paper**

(i) The "Hybrid Photon" model eliminates the need for the ad hoc postulate, "wave-particle duality". Interpretation of Quantum Mechanics becomes stronger and more realistic without this ad hoc postulate.

(ii) Superposition Effect can be purely classical or quantum mechanical, depending upon the situation. Semi-classical model explains both situations. Our current technology cannot directly measure a single photon of energy $\sim 4 \times 10^{-18}$ $Joules$ (visible range). We detect current pulse containing, perhaps, billions of amplified electrons. "Clicks" of billions of electrons should not be confused with imaginary single photon.

(iii) Mathematically, the linear superposition principle (SP) represents linear sum of two or more complex-amplitudes stimulation induced on **material dipoles** of some "detector" with polarizability $\chi$ (could be the boundary layer of a beam splitter or a quantum photodetector).

$$A_{res.} = \chi a_1 e^{i2\pi\nu t} + \chi a_2 e^{i2\pi\nu(t+\tau)}$$

We must not directly sum the wave amplitudes. Non-Interaction of Waves (NIW) is fundamental characteristic of all propagating waves. Besides, a single stable elementary particle cannot simultaneously carry multiple physical values for the same physical parameter. The above superposition equation contain too many real physical parameters to be carried by a **single and stable elementary particle**. "Single particle interference" is an impossible demand by our causal physical equation that works.

(iv) Further, the observable Superposition Effect is a non-linear quadratic process. Yhe observable effect must be mediated by detecting entity with the capability to carry out such nonlinear quadratic process after receiving two or more simultaneous stimulating signals. Detector absorbs signal from both stimulating entities as is evident from the presence of two different "amplitude" factors.

$$|A_{res.}|^2 = \chi^2 a_1^2 + \chi^2 a_2^2 + 2\chi^2 a_1 a_2 \cos 2\pi\nu\tau$$
$$= 2\chi^2 a^2[1 + \cos 2\pi\nu\tau]; \text{ only when } a_1 = a_2 \equiv a$$